\documentclass[a4paper,fleqn,usenatbib]{mnras}
\usepackage{epsfig}   
\usepackage{amsmath}
\usepackage{mathptmx}
\usepackage{txfonts}
\usepackage[T1]{fontenc}
\usepackage{ae,aecompl}
\usepackage{graphicx}	
\usepackage{amssymb}	
\usepackage{scalerel}

\newcommand{\be}{\begin{equation}}
\newcommand{\e}{\end{equation}}
\newcommand{\bear}{\begin{eqnarray}}
\newcommand{\ear}{\end{eqnarray}}

\def\aj{AJ}
\def\apj{ApJ}
\def\apjs{ApJS}

\def\mnras{MNRAS}
\def\aap{A\&A}

\def\apjs{ApJS}
\def\apjl{ApJ Letters}

\title[Fuzzifying galaxy colours] {A method for classification of red,
  blue and green galaxies using fuzzy set theory}
    
\author[Pandey, B.] {Biswajit Pandey\thanks{E-mail:
    biswap@visva-bharati.ac.in} \\ Department of Physics,
  Visva-Bharati University, Santiniketan, Birbhum, 731235, India\\ }
   
 \date{\today}

 \pubyear{2020}
  
\begin{document}
\label{firstpage}
\pagerange{\pageref{firstpage}--\pageref{lastpage}}      
\maketitle
       
\begin{abstract}
Red and blue galaxies are traditionally classified using some specific
cuts in colour or other galaxy properties, which are supported by
empirical arguments. The vagueness associated with such cuts are
likely to introduce a significant contamination in these
samples. Fuzzy sets are vague boundary sets which can efficiently
capture the classification uncertainty in the absence of any precise
boundary. We propose a method for classification of galaxies according
to their colours using fuzzy set theory. We use data from the SDSS to
construct a fuzzy set for red galaxies with its members having
different degrees of `redness'. We show that the fuzzy sets for the
blue and green galaxies can be obtained from it using different fuzzy
operations. We also explore the possibility of using fuzzy relation to
study the relationship between different galaxy properties and discuss
its strengths and limitations.
\end{abstract}


\section{Introduction}
Colour is considered to be one of the fundamental properties of a
galaxy. It is defined as the ratio of fluxes in two different bands
and characterizes the stellar population of a galaxy. Understanding
the distribution of galaxy colours and their evolution can provide
important clues to galaxy formation and evolution.

The modern redshift surveys like SDSS \citep{york} have now measured
the photometric and spectroscopic information of a large number of
galaxies. Using SDSS, \citet{strateva} first revealed a striking
bimodal feature in the distribution of galaxy colours.  Subsequent
studies with SDSS and other surveys
\citep{blanton03,bell1,baldry04} confirmed the colour bimodality
and quantified it in greater detail. The observed colour bimodality
indicates the existence of two different populations namely the `red
sequence' and `blue sequence' with a significant overlap between
them. The overlapping region is often termed as `green
valley'. \citet{driver} argued that galaxy colours are an outcome of
the mixture of colours from stars in the disc and bulge of a galaxy
and the colour bimodality can be related to the bimodal distribution
of their bulge to disc mass ratios. A number of studies show that
colour bimodality is sensitive to luminosity, stellar mass and
environment \citep{balogh, baldry06}.

The origin of the observed colour bimodality must be explained by
successful models of galaxy formation. Considerable efforts have been
directed towards reproducing the observed distribution of galaxy
colours using different semi-analytic models of galaxy formation
\citep{menci,driver,cattaneo1,cattaneo2,cameron,trayford,nelson,correa19}.
Significant number of studies have been also devoted to understand the
spatial distribution of red and blue galaxies using various clustering
measures like two-point correlation function \citep{zehavi11},
three-point correlation function \citep{kayo}, genus \citep{hoyle1},
filamentarity \citep{pandey06} and local dimension
\citep{pandey20}. The studies of mass function of red and blue
galaxies \citep{drory, taylor} play a crucial role in guiding the
theories of galaxy formation and evolution. Any such analysis requires
one to classify the red and blue galaxies using some operational
definition. But colour is a subjective judgment and it is difficult to
classify galaxies according to their colours in an objective manner.
Also one should remember that it is the observed bimodal distribution
of galaxy colours that motivates us to classify the galaxies according
to their colours. In reality, no galaxies can be regarded as truly
either `red' or `blue' based on their colours.  \citet{strateva}
prescribed that the red and blue galaxies can be separated by using an
optimal colour separator $(u-r)=2.22$.  However a substantial overlap
between the two populations prohibit us to objectively define galaxies
as `red' and `blue' based on their colours. The boundary separating
the two populations remains arbitrary and currently there are no
meaningful theoretical argument favoring any specific cut over
others. Applying a hard-cut to define samples of red and blue galaxies
is expected to introduce significant contamination in these
samples. Previously, \citet{coppa} use an unsupervised fuzzy partition
clustering algorithm applied on the principal components of a PCA
analysis to study the bimodality in galaxy colours. Another study by
\citet{norman} use Bayesian statistical methods instead of sharp cuts
in parameter space for such classification problems.

The fuzzy set theory was first introduced by Lotfi A. Zadeh
\citep{zadeh} which has been applied in various fields such as
industrial automation \citep{zadeh1}, control system \citep{lugli},
image processing \citep{rosenfeld1}, pattern recognition
\citep{rosenfeld2} and robotics \citep{wakileh}. It provides a natural
language to express vagueness or uncertainty associated with imprecise
boundary.

In this Letter, we propose a framework to describe galaxy colours
using a fuzzy set theoretic approach. We also study the prospects of
using fuzzy relations to understand the relationship between different
galaxy properties by treating them as fuzzy variables.

The Letter is organized as follows: We describe the SDSS data in
Section 2, explain the method in Section 3 and present the results and
conclusions in Section 4.

\section{SDSS DATA}
The Sloan Digital Sky Survey (SDSS) is one of the most successful sky
surveys of modern times. The SDSS has so far measured the photometric
and spectroscopic information of more than one million of galaxies and
quasars in five different bands. We use data from SDSS DR16
\citep{ahumada} for the present analysis. We retrieve data from the
SDSS SkyServer \footnote{https://skyserver.sdss.org/casjobs/} using
Structured Query Language. We download the spectroscopic information
of all the galaxies with r-band Petrosian magnitude $ 13.5 \le r_{p} <
17.77$ and located within $135^{\circ} \leq \alpha \leq 225^{\circ}$,
$0^{\circ} \leq \delta \leq 60^{\circ}$, $z < 0.3$.  Here $\alpha$,
$\delta$ and $z$ are right ascension, declination and redshift
respectively. We obtain a total $376495$ galaxies which satisfy these
cuts. We then apply a cut to the K-corrected and extinction corrected
$r$-band absolute magnitude $-21 \ge M_r \ge -23 $ which corresponds to a
redshift cut of $0.041 \le z \le 0.120$. This produces a volume limited
sample with $103984$ galaxies. We use a $\Lambda$CDM cosmological
model with $\Omega_{m0}=0.315$, $\Omega_{\Lambda0}=0.685$ and
$h=0.674$ \citep{planck18}.

\section{Method of Analysis}
\subsection{Fuzzy set}
A fuzzy set $A$ in an Universal set $X$ is defined as a set of ordered
pairs \citep{zadeh},
\begin{eqnarray}
A=\big{\{}\,(\,x,\,\mu_{\scaleto{A}{3.5pt}}(x)\,) \,\,|\,\,x \in X\,\big{\}} 
\label{eq:fuzzy1}
\end{eqnarray}
where $\mu_{\scaleto{A}{3.5pt}}(x)$ is the membership function of $x$
in $A$. $\mu_{\scaleto{A}{3.5pt}}(x)$ maps elements of the Universal
crisp set $X$ into real numbers in $[0,1]$
i.e. $\mu_{\scaleto{A}{3.5pt}}(x): X \to [0,1]$. It provides the
degree of membership of $x$ in $X$.  Thus fuzzy sets are vague
boundary sets. The maximum value of the membership degree
characterizes the height of a fuzzy set. Fuzzy sets with height one
are known as normal fuzzy sets. The membership function of a fuzzy set
can have different shapes such as: triangular, trapezoidal, Gaussian,
sigmoidal etc. in different contexts.

\subsection{Fuzzy relation}
A fuzzy relation $S$ between two fuzzy sets $A_1$ and $A_2$ in product
space $X \times Y$ is a mapping from the product space to an interval
$[0,1]$ and is defined as,
\begin{eqnarray}
S=\big{\{}\,(x,y) \,\,|\,\, x \in A_1, y \in A_2\,\big{\}}
\label{eq:fuzzy4}
\end{eqnarray}
$S$ describes the relations between the elements of the two fuzzy sets
$A_1$ and $A_2$. The strength of the relation between the ordered
pairs of the two sets is expressed by the membership function of the
relation $\mu_{\scaleto{A_1 \times
    A_2}{3.5pt}}(x,y)=\mu_{\scaleto{S}{3.5pt}}(x,y)=min\big{\{}\,\mu_{\scaleto{A_1}{3.5pt}}(x),\,\mu_{\scaleto{A_2}{3.5pt}}(y)\,\big{\}}$.
The fuzzy relation $S$ itself is a fuzzy set whose membership function
is represented with a matrix known as relation matrix.

\begin{figure*}
\resizebox{6.0cm}{!}{\rotatebox{0}{\includegraphics{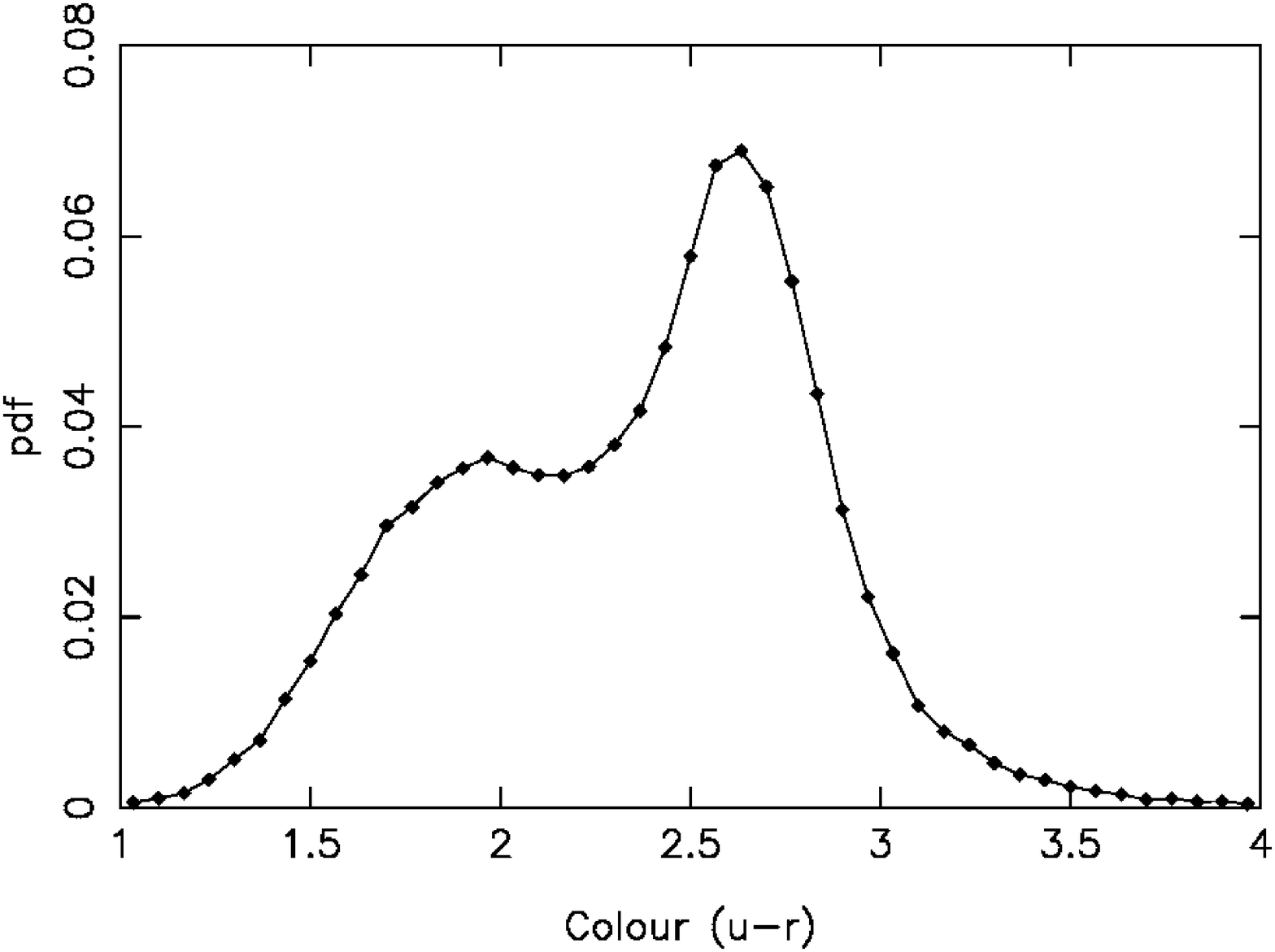}}}%
\resizebox{6.0cm}{!}{\rotatebox{0}{\includegraphics{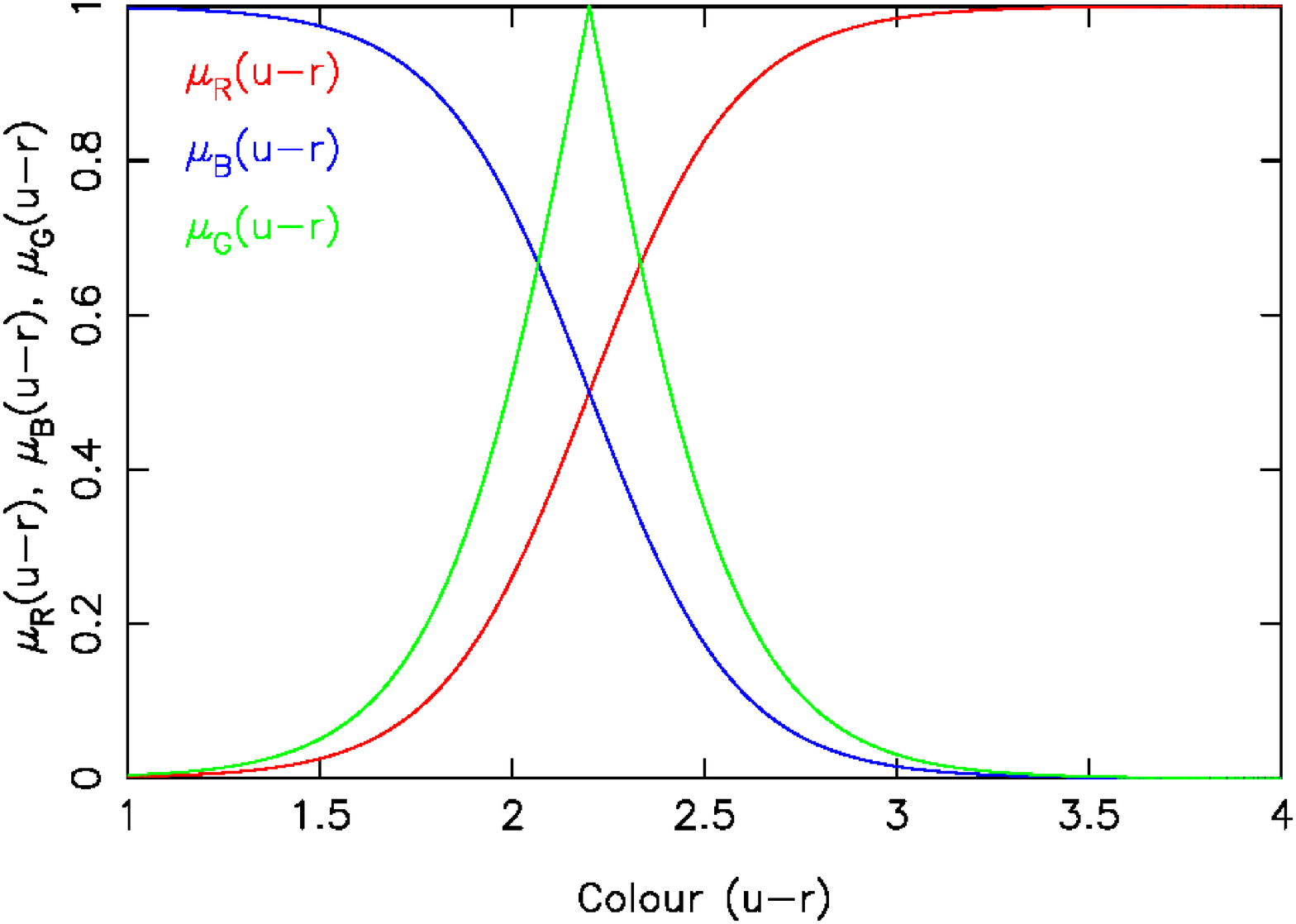}}}\\
\resizebox{6.0cm}{!}{\rotatebox{0}{\includegraphics{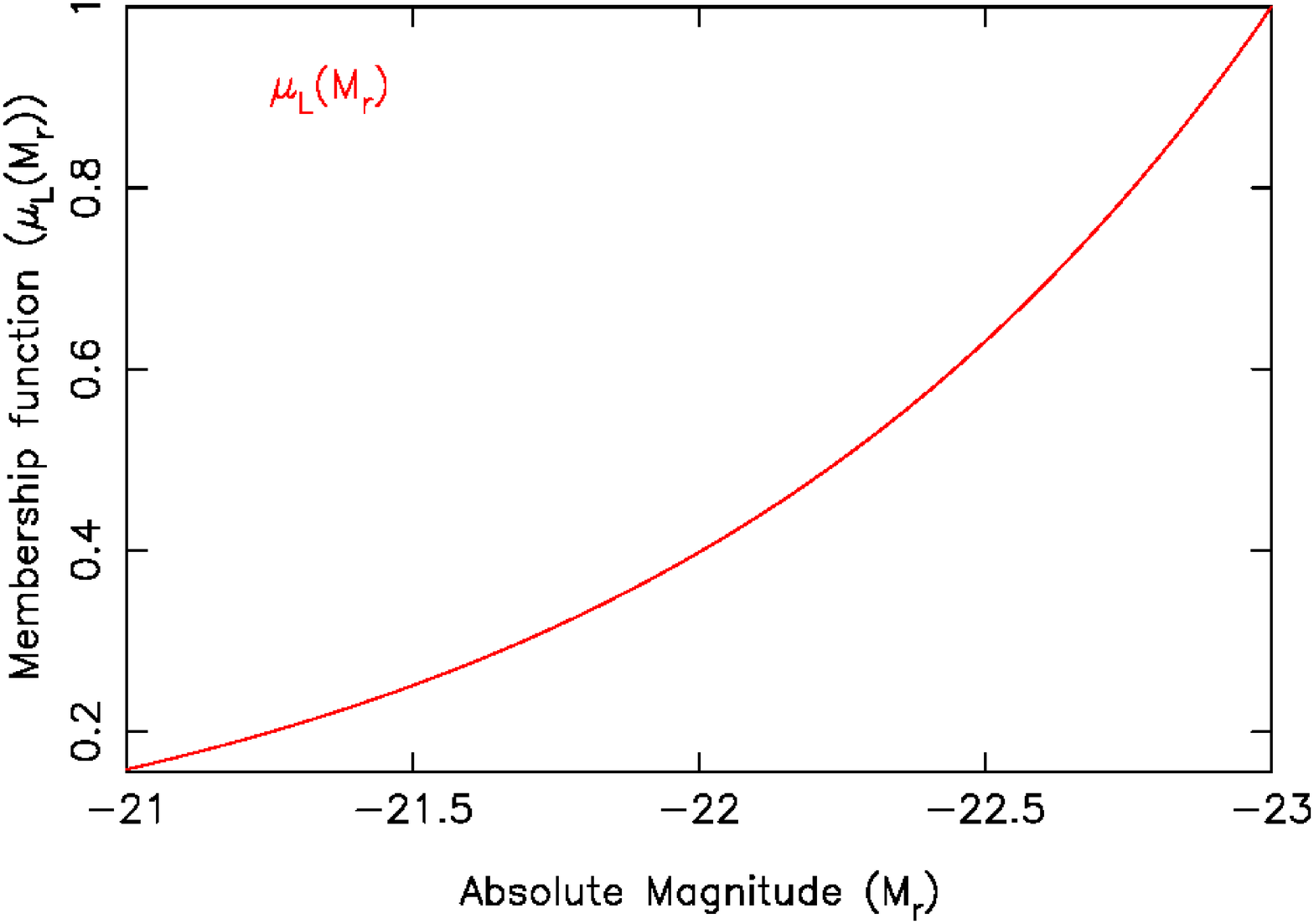}}}%
\resizebox{6.0cm}{!}{\rotatebox{0}{\includegraphics{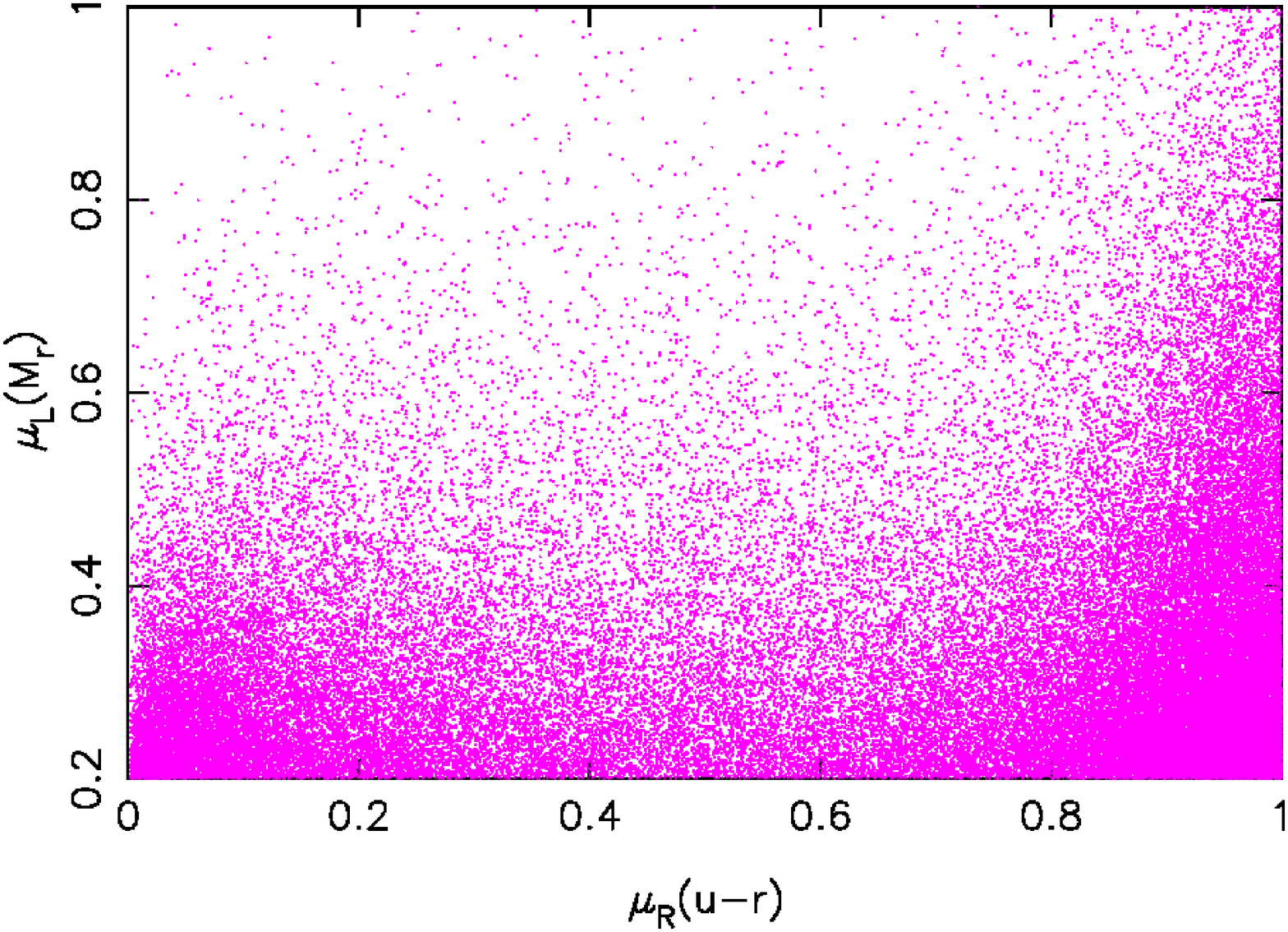}}}\\
\caption{The top left panel of this figure shows the pdf of the
  $(u-r)$ colour of SDSS galaxies. The top right panel shows the
  membership functions
  $\mu_{\scaleto{R}{3.5pt}}({\scaleto{u-r}{3.5pt}})$,
  $\mu_{\scaleto{B}{3.5pt}}({\scaleto{u-r}{3.5pt}})$,
  $\mu_{\scaleto{g}{3.5pt}}({\scaleto{u-r}{3.5pt}})$ of the fuzzy sets
  corresponding to red, blue and green galaxies in the SDSS as a
  function $(u-r)$ colour. The bottom left panel shows the membership
  function $\mu_{\scaleto{L}{3.5pt}}$ of the fuzzy set for luminosity
  of the SDSS galaxies as a function of r-band absolute magnitude
  ($M_r$). The membership function
  $\mu_{\scaleto{L}{3.5pt}}({\scaleto{M_r}{3.5pt}})$ for luminosity
  and membership function
  $\mu_{\scaleto{R}{3.5pt}}({\scaleto{u-r}{3.5pt}})$ for red galaxies
  are plotted against each other in the bottom right panel.}
\label{fig:one}
\end{figure*}

\begin{figure*}
\resizebox{9cm}{!}{\rotatebox{0}{\includegraphics{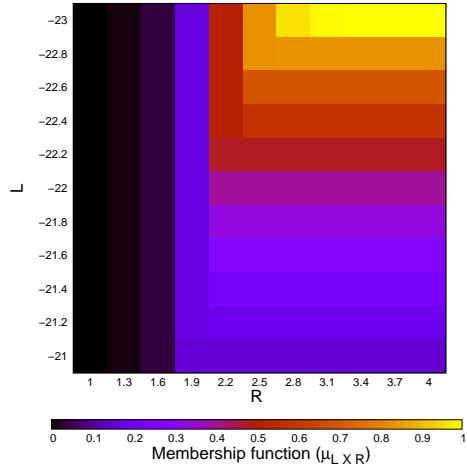}}}%
\caption{This shows the fuzzy relation between the fuzzy sets for
  luminosity ($L$) and red galaxies ($R$).}
  \label{fig:two}
\end{figure*}

\subsection{Definition of fuzzy sets for different colours and luminosity}
Application of a hard-cut yields two crisp sets corresponding to red
and blue galaxies. This denies a gradual transition between the two
populations which is mathematically correct but unrealistic. In
reality, there is an uncertainty involved in the decision of labelling
a galaxy as either `red' or `blue' whenever the colour falls in the
neighbourhood of the precisely defined border. This uncertainty is
completely ignored by the hard-cut separator. In fuzzy set theory, an
element may partly belong to multiple fuzzy sets with different degree
of memberships thereby allowing to capture this uncertainty whenever
the boundaries are imprecise.

We do not label galaxies as purely red, blue or green. Rather we
attach `redness', `blueness' or `greenness' to all the galaxies. We
use SDSS data to define fuzzy sets corresponding to three colours red,
blue and green. We define a fuzzy set $R$ corresponding to `redness'
of galaxies using $(u-r)$ colour of $103984$ galaxies in the volume
limited sample as,
\begin{eqnarray}
R=\big{\{}\, (u-r,\, \mu_{\scaleto{R}{3.5pt}}({\scaleto{u-r}{3.5pt}}))\,\,|\,\, (u-r) \in X \,\big{\}} 
\label{eq:fuzzy5}
\end{eqnarray}
Here $X$ is the Universal set of $(u-r)$ colour of all galaxies. We
choose a sigmoidal membership function,
\begin{eqnarray}
\mu_{\scaleto{R}{3.5pt}}(u-r;a,c)=\frac{1}{1+e^{-a[(u-r)-c]}}
\label{eq:fuzzy6}
\end{eqnarray}
where $a$ and $c$ are constants. We choose $a=5.2$ and $c=2.2$ for our
analysis. The choice of the sigmoidal membership function is based on
the fact that galaxies with larger $(u-r)$ colour are considered to be
progressively redder and those with a very high $(u-r)$ colour can be
reliably tagged as `mostly red'. One may choose a linear shape for it
but non-linear functions are known to perform better in most practical
situations \citep{kwanglee}. It may be noted that the parameter $c$
represents the crossover point of the fuzzy set $R$ where
$\mu_{\scaleto{R}{3.5pt}}({\scaleto{u-r}{3.5pt}})=0.5$ and parameter
$a$ denotes the slope at the crossover point. The crossover point of a
fuzzy set is the location where the fuzzy set has maximum uncertainty
or vagueness. SDSS observations of the bimodal distribution of $(u-r)$
colour show that the two peaks corresponding to `red' and `blue'
population merge together at $u-r\sim2.2$. It is most difficult to
classify a galaxy as red or blue around this $(u-r)$ colour. The slope
$a$ at the cross-over point is somewhat arbitrarily chosen in this
analysis where we have ensured that the galaxy with highest and lowest
$(u-r)$ colour have
$\mu_{\scaleto{R}{3.5pt}}({\scaleto{u-r}{3.5pt}})=1$ and
$\mu_{\scaleto{R}{3.5pt}}({\scaleto{u-r}{3.5pt}})=0$ respectively. We
believe that the choice of the parameter $a$ can be improved by
training an artificial neural network with sample data.

A larger `redness' is associated with smaller `blueness' and vice
versa. So we can define a fuzzy set $B$ corresponding to `blueness' of
galaxies by simply taking a fuzzy complement of the normal fuzzy set
$R$ i.e. $B=R^{c}$. The membership function
$\mu_{\scaleto{B}{3.5pt}}({\scaleto{u-r}{3.5pt}})$ of fuzzy set $B$
can be simply obtained as follows,
\begin{eqnarray}
\mu_{\scaleto{B}{3.5pt}}({\scaleto{u-r}{3.5pt}})=1-\mu_{\scaleto{R}{3.5pt}}({\scaleto{u-r}{3.5pt}}), \,  \forall (u-r) \in X
\label{eq:fuzzy7}
\end{eqnarray}

We define the green galaxies as those which simultaneously belong to
both the fuzzy sets $R$ and $B$. The fuzzy set $G$ corresponding to
the `greenness' of galaxies is defined from the two fuzzy sets $R$ and
$B$ by taking a fuzzy intersection between them i.e. $G=R \cap B$. The
fuzzy set $G$ is pointwise defined by the membership function
$\mu_{\scaleto{G}{3.5pt}}({\scaleto{u-r}{3.5pt}})$ as
\begin{eqnarray}
\mu_{\scaleto{G}{3.5pt}}({\scaleto{u-r}{3.5pt}})= 2 \, min\big{\{}\,
\mu_{\scaleto{R}{3.5pt}}({\scaleto{u-r}{3.5pt}}),\,
\mu_{\scaleto{B}{3.5pt}}({\scaleto{u-r}{3.5pt}})\,\big{\}},\, \forall
(u-r) \in X
\label{eq:fuzzy8}
\end{eqnarray}
where $min$ is the minimum operator and the prefactor of 2 is used for
normalization. The intersection of the fuzzy sets $R$ and $B$ will be
a subnormal fuzzy set with a height $0.5$ at the crossover point
$(u-r)=2.2$. However a galaxy with $(u-r)$ colour of $2.2$ should be
maximally green with a membership function
$\mu_{\scaleto{G}{3.5pt}}({\scaleto{u-r}{5.5pt}})=1$.

It may be noted here that the membership functions do not provide the
probability of a galaxy being `red', `blue' or `green'. The membership
functions show the possibility of being contained in a particular
fuzzy set. They are not similar to the likelihood functions in
probability theory where the overall likelihood function need to be
normalized to unity.  The fuzzy set theory is based on the idea of
possibility rather than probability. Both probability theory and fuzzy
theory express uncertainty and have similarities in certain
aspects. But there are some important differences between the two. For
example the sum of the probabilities in a probability distribution
must be equal to $1$. Contrary to this, there is no such limit for the
sum in a possibility distribution defined on an Universal set
\citep{kwanglee}.

It is worthwhile to mention here that the hard-cut samples can be
easily obtained from these fuzzy sets by constructing appropriate
$\alpha-$level sets which are crisp sets. The elements that belong to
a fuzzy set with at least a degree of membership $\alpha$ is called
the $\alpha$-level set:
\begin{eqnarray}
A_{\alpha}=\big{\{}\, x \in X \,\,| \,\, \mu_{\scaleto{A}{3.5pt}}(x) \geq \alpha\,\big{\}}
\label{eq:fuzzy9}
\end{eqnarray}
An $\alpha$-level set selects the objects above a certain degree of
possibility. A higher alpha-cut to a fuzzy set implies that the
members of the resulting crisp set belong to a particular class with
higher possibility.

We also define a fuzzy set for luminosity of SDSS galaxies as,
\begin{eqnarray}
L=\big{\{}\, (\,M_r,\,\mu_{\scaleto{L}{3.5pt}}({\scaleto{M_r}{4.5pt}}))\,\,|\,\,M_r \in Y \,\big{\}} 
\label{eq:fuzzy10}
\end{eqnarray}
where $Y$ is the Universal set of r-band absolute magnitude ($M_r$) of
all galaxies. The membership function of this fuzzy set is,
\begin{eqnarray}
\mu_{\scaleto{L}{3.5pt}}({\scaleto{M_r}{4.5pt}})=\frac{1}{(2.512)^{[M_r-(M_r)_{brightest}]}}\,,\,\forall M_r \in Y
\label{eq:fuzzy11}
\end{eqnarray}
The choice of this membership function is based on the fact that a
difference of $1$ in absolute magnitude corresponds to a factor of
$2.512$ in luminosity. Thus the brightest galaxy in the sample have a
membership function of $1$ and the membership function
$\mu_{\scaleto{L}{3.5pt}}({\scaleto{M_r}{4.5pt}})$ for the galaxies in
$L$ are assigned following \ref{eq:fuzzy11}. It should be noted that
the fuzzy set $L$ is not defined to categorize galaxies according to
their luminosity. We construct the fuzzy set $L$ in order to study the
fuzzy relation between it and the `red' fuzzy set.

\section{RESULTS AND CONCLUSIONS}
In the top left panel of \autoref{fig:one}, we show the pdf of $(u-r)$
colour of SDSS galaxies in the volume limited sample analyzed
here. The bimodal nature of the $(u-r)$ colour can be clearly seen in
this figure. It has been shown that the colour distribution can be
well described by a double Gaussian \citep{balogh, baldry06,
  taylor}. We observe that the pdf of the $(u-r)$ colour exhibit two
distinct peaks which merge together at $(u-r)\sim 2.2$. The $(u-r)$
colour of SDSS galaxies show a gradual transition on either sides of
this merging point. Traditionally, the galaxies are classified as
`red' or `blue' by employing a hard-cut around the merging point of
the two peaks. However, any $(u-r)$ colour close to this border should
be regarded as an equal evidence for both the classes. The uncertainty
reaches its maximum at the border which is completely overlooked by
such a hard-cut separator. We treat this merging point as the point of
maximum confusion and use it as the crossover point
($\mu_{\scaleto{R}{3.5pt}}({\scaleto{u-r}{3.5pt}})=0.5$) for the
membership function (\autoref{eq:fuzzy6}) of the fuzzy set of red
galaxies ($R$) in SDSS. We obtain the fuzzy set of blue galaxies ($B$)
in the SDSS by taking a fuzzy complement of the fuzzy set $R$. The
green galaxies belong to an intermediate class which are most
difficult to disentangle. We carry out a fuzzy intersection between
the two fuzzy sets $R$ and $B$ to construct the fuzzy set of green
galaxies ($G$) in the SDSS. We compute the membership functions of all
the SDSS galaxies in these fuzzy sets following \autoref{eq:fuzzy6},
\autoref{eq:fuzzy7} and \autoref{eq:fuzzy8}. The membership functions
$\mu_{\scaleto{R}{3.5pt}}({\scaleto{u-r}{3.5pt}})$,
$\mu_{\scaleto{B}{3.5pt}}({\scaleto{u-r}{3.5pt}})$ and
$\mu_{\scaleto{G}{3.5pt}}({\scaleto{u-r}{3.5pt}})$ corresponding to
red, blue and green galaxies respectively are shown as a function of
$(u-r)$ colour of SDSS galaxies in the top right panel of
\autoref{fig:one}. We can see that galaxies are maximally green at the
merging point of the peaks where classifying a galaxy as red or blue
is most uncertain.

We define the fuzzy set $L$ for luminosity of SDSS galaxies using
their r-band absolute magnitudes. The membership function
$\mu_{\scaleto{L}{3.5pt}}({\scaleto{M_r}{5pt}})$ of the fuzzy set $L$
is defined according to variations in luminosity with absolute
magnitude (\autoref{eq:fuzzy11}). We show the membership function
$\mu_{\scaleto{L}{3.5pt}}({\scaleto{M_r}{5pt}})$ of the fuzzy set $L$
for SDSS galaxies as a function of r-band absolute magnitude ($M_r$)
in the bottom left panel of \autoref{fig:one}.  We also plot the
membership function ($\mu_{\scaleto{L}{3.5pt}}({\scaleto{M_r}{5pt}})$)
of set $L$ against the membership function
$\mu_{\scaleto{R}{3.5pt}}({\scaleto{u-r}{3.5pt}})$ of set $R$ of the
SDSS galaxies in the bottom right panel of \autoref{fig:one}.  The
bottom right panel of \autoref{fig:one} shows that there are two
distinct peaks located near the both extremities of
$\mu_{\scaleto{R}{3.5pt}}({\scaleto{u-r}{3.5pt}})$ around which the
members of the two fuzzy sets are clustered. It shows that the bright
galaxies are preferentially red and then blue but they have least
possibility of being green. The observed trend is somewhat more
conspicuous than those observed in the color-magnitude diagram. Unlike
the color-magnitude diagram, it shows the relationship between the two
membership functions which themselves express the degree of `redness'
and `brightness'. A combination of two $\alpha$-cuts within $0$ to $1$
on the two membership functions
$\mu_{\scaleto{R}{3.5pt}}({\scaleto{u-r}{5.5pt}})$ and
$\mu_{\scaleto{L}{3.5pt}}({\scaleto{M_r}{6.5pt}})$ can be used to
define a galaxy sample with desired properties. This can be extended
to any number of fuzzy sets and hence allows a greater flexibility in
classifying galaxies with different properties.

We compute the fuzzy relation $S$ between the fuzzy sets $L$ and $R$
using \autoref{eq:fuzzy4}. The resulting relation matrix is shown in
\autoref{fig:two}. This shows the strengths of association for
different possible combinations of r-band absolute magnitude ($M_r$)
and colour ($u-r$). We find that for $(u-r)>2.2$, the membership
function of the fuzzy relation gradually increases with increasing
luminosity and the association is strongest near the largest
luminosity.  Contrary to this, no such gradients are seen for
$(u-r)<2.2$. This implies that brighter galaxies are fairly redder for
galaxies with $(u-r)$ colour larger than $2.2$. But this is not
necessarily the case when we consider the region below $(u-r)<2.2$
which represents less redder i.e. bluer galaxies. However, one should
remember that such a boundary is not precise and a mixed behaviour can
be observed in the proximity of $(u-r)=2.2$. In probability theory, we
may be simply interested in calculating the covariance of two random
variables $M_r$ and $(u-r)$ which can be represented by a $2 \times 2$
matrix. The elements corresponding to each variables in that case are
treated as crisp as there are no uncertainty about their
memberships. Also the covariance matrix does not provide the
association between all possible elements of two sets. The fuzzy
relation delineate the relationship between two fuzzy sets by
measuring the association between all possible elements of two
sets. This allows one to distinguish the degree of association between
the two variables across the different regions of parameter space.

Training an artificial neural network with sample data to learn about
the membership function may help a more efficient construction of the
membership function. Further, we are using Type-1 fuzzy sets whose
membership functions are crisp sets. So there are no uncertainties in
the membership functions of the fuzzy sets defined here. One can
include the uncertainties in the membership functions by using Type-2
fuzzy sets. A Type-2 fuzzy set is a fuzzy set whose membership
function is a Type-1 fuzzy set. We plan to address these issues in a
future work. The analysis presented in this work is to be considered
as a first step towards a more complete and rigorous treatment.

Finally, we note that fuzzy set theory has great potential to become
an useful tool in highly data driven fields of astrophysics and
cosmology. An imprecise boundary between different class of objects
and their properties are quite common in astronomical datasets. In
cosmology, reproducing different observed galaxy properties using
various semi-analytic models help us to fine-tune our understanding of
galaxy formation and evolution. There are complex interplay between
various factors (e.g. density, environment) and galaxy properties
which leads to large scatter in their values and relationships with
each other. Fuzzy relations and their compositions in such situations
may help us understand their relationships from a different
perspective and can serve as a complementary measure to other existing
tools.
\section{ACKNOWLEDGEMENT}
The author thanks Didier Fraix-Burnet for some useful comments and
suggestions. I would like to thank the SDSS team for making the data
public. I would also like to thank Suman Sarkar for the help with the
SDSS data. A financial support from the SERB, DST, Government of India
through the project CRG/2019/001110 is duly acknowledged. The author
acknowledges IUCAA, Pune for providing support through associateship
programme. 

Funding for the SDSS and SDSS-II has been provided by the Alfred
P. Sloan Foundation, the Participating Institutions, the National
Science Foundation, the U.S. Department of Energy, the National
Aeronautics and Space Administration, the Japanese Monbukagakusho, the
Max Planck Society, and the Higher Education Funding Council for
England. The SDSS Web Site is http://www.sdss.org/.

The SDSS is managed by the Astrophysical Research Consortium for the
Participating Institutions. The Participating Institutions are the
American Museum of Natural History, Astrophysical Institute Potsdam,
University of Basel, University of Cambridge, Case Western Reserve
University, University of Chicago, Drexel University, Fermilab, the
Institute for Advanced Study, the Japan Participation Group, Johns
Hopkins University, the Joint Institute for Nuclear Astrophysics, the
Kavli Institute for Particle Astrophysics and Cosmology, the Korean
Scientist Group, the Chinese Academy of Sciences (LAMOST), Los Alamos
National Laboratory, the Max-Planck-Institute for Astronomy (MPIA),
the Max-Planck-Institute for Astrophysics (MPA), New Mexico State
University, Ohio State University, University of Pittsburgh,
University of Portsmouth, Princeton University, the United States
Naval Observatory, and the University of Washington.

\section{Data Availability}
The data underlying this article are available in
https://skyserver.sdss.org/casjobs/. The datasets were derived from
sources in the public domain: https://www.sdss.org/

\bsp	
\label{lastpage}
\end{document}